\documentclass{iaus}
\usepackage{graphicx}

\title[Early-type stars in the Magellanic Clouds]
{The Properties of Early-type Stars in the Magellanic Clouds}

\author[C. J. Evans]{Christopher J. Evans$^1$}
\affiliation{$^1$UK Astronomy Technology Centre, Royal Observatory Edinburgh, 
Blackford Hill, Edinburgh, EH9 3HJ, UK\\ email: {\tt cje@roe.ac.uk} \\[\affilskip]}

\pubyear{2008}
\volume{256}  
\pagerange{1--12}
\jname{The Magellanic System: Stars, Gas, and Galaxies}
\editors{Jacco Th. van Loon \& Joana M. Oliveira, eds.}
\begin{document}

\maketitle

\begin{abstract}
The past decade has witnessed impressive progress in our understanding
of the physical properties of massive stars in the Magellanic Clouds,
and how they compare to their cousins in the Galaxy.  I summarise new
results in this field, including evidence for reduced mass-loss
rates and faster stellar rotational velocities in the Clouds, and
their present-day compositions.  I also discuss the stellar
temperature scale, emphasizing its dependence on metallicity across
the entire upper-part of the Hertzsprung-Russell diagram.
\end{abstract}

\firstsection 
\section{Introduction}

The prime motivation for studies of early-type stars in the Magellanic
Clouds over the past decade has been to quantify the effect of
metallicity ($Z$) on their evolution.  The intense out-flowing winds
in massive stars are thought to be driven by momentum transferred from
the radiation field to metallic ions (principally iron) in their
atmospheres; a logical consequence of this mechanism is that the wind
intensities should vary with $Z$ (Kudritzki et al., 1987).  Monte
Carlo simulations predict that, for stars with T$_{\rm
eff}\,>$\,25,000\,K, the wind mass-loss rates should scale with
metallicity as $Z^{0.69}$ (Vink et al., 2000; 2001).  This has a
dramatic impact on their subsequent evolution.  For example, an O-type star in
the SMC should lose significantly less mass over its lifetime than a
star in the Galaxy, thus retaining greater angular momentum.  This
could then lead to different late-phases of evolution such as the
type of core-collapse supernova (SN),
and offers a potential channel for long duration, gamma-ray bursts
at low $Z$.

The $Z$-dependence of the {\em initial} rotational velocity
distributions of massive stars and the importance of
rotationally-induced mixing have also been active areas of research.
For instance, \cite{mgm99} noted that the relative fraction of Be- to
B-type stars increases with decreasing metallicity\footnote{The sample
of Maeder et al. comprised only one SMC cluster, NGC\,330, long known
to have a significant Be-fraction (e.g. Grebel et al., 1992) and
sometimes suggested as a `pathological' case.  However, new results
from Martayan et al. (these proceedings) also find similarly large
fractions for other SMC clusters.}, suggesting that this might arise
from faster rotational velocities at lower $Z$.  The recent generation
of evolutionary models has explored the effects of rotational mixing
(e.g. Heger \& Langer, 2000; Meynet \& Meynet, 2000), with the
prediction of larger relative surface-nitrogen enhancements at faster
rotation rates, and at lower Z (Maeder \& Meynet, 2001).

To date, we have lacked sufficient observations to explore the
effects of metallicity thoroughly.  Robust empirical results were
needed with which to confront both the stellar wind and evolutionary
models for early-type stars -- here I summarise recent observations
and quantitative analyses toward this objective.

\newpage
\section{Unique insights from ultraviolet observations}

Ultraviolet (UV) spectroscopy provides an invaluable complement to
optical spectroscopy in determination of the physical parameters of
early-type stars.  The terminal velocity ($v_\infty$) of the stellar
wind can be measured from the saturated cores of the resonance lines,
with information on the stratification and ionization of the wind
provided by comparisons with model atmospheres.  Unfortunately, only a
handful of the brightest OB-type stars in the Clouds were within reach
of high-dispersion observations with the {\em International
Ultraviolet Explorer (IUE)}.

The {\em Hubble Space Telescope (HST)} and the {\em Far Ultraviolet
Spectroscopic Explorer (FUSE)} have provided essential observations of the 
winds of massive stars in the Clouds.  Although limited by relatively
low spectral resolution, the {\em HST} Faint Object Spectrograph (FOS)
was used in the early 1990s to observe tens of stars in the Clouds.
These spectra provided morphological evidence of weaker stellar winds
in the SMC when compared to Galactic standards (Walborn et al., 1995),
as well as estimates of $v_\infty$ (e.g., Prinja \& Crowther, 1998).
More recently, the {\em HST}\/ Space Telescope Imaging Spectrograph
(STIS) and {\em FUSE} have both delivered spectra at resolutions of
$>$10$^4$, sufficient to resolve interstellar features clearly and to
enable detailed comparisons with synthetic spectra.

\begin{figure}[h]
\vspace*{-0.35 cm}
\begin{center}
 \includegraphics[width=4.75in]{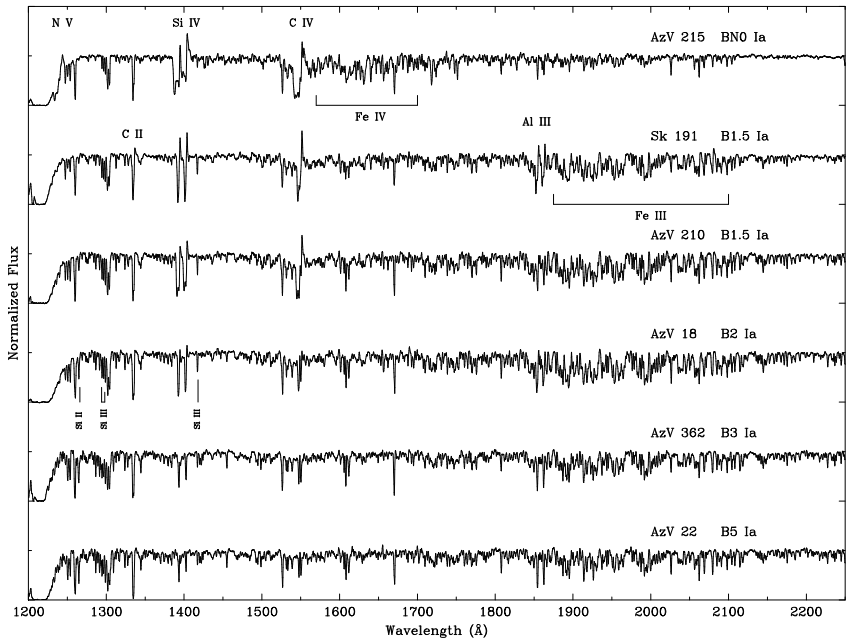} 
 \vspace*{-0.1 cm}
 \caption{STIS spectra of B-type supergiants in the SMC (Evans et al., 2004a).  Note the 
distinctive (although weak) P~Cygni emission in the N~{\small V}, Si~{\small IV} and C~{\small IV}
lines in AzV~215, and the iron `forests' which neatly illustrate the change in the predominant
ionization stage in the early B-type domain.}\label{fig1}
\end{center}
\end{figure}

STIS spectroscopy of 29 OB-type SMC stars was presented by \cite{wal00} and 
\cite{elw04}, providing further morphological evidence for weaker wind features in
SMC stars when compared to their Galactic counterparts.  While the
intensity of the P~Cygni emission is reduced in B-type supergiants in
the SMC (Fig.~\ref{fig1}), the terminal velocities ($v_\infty$) are
not significantly slower (Evans et al., 2004b), consistent with the
weak metallicity-dependence predicted by theory, $v_\infty \propto
Z^{0.13}$ (Leitherer et al., 1992); in the O-type domain this relation
manifests itself more clearly, e.g. Walborn et al.  (1995).

The ratio of terminal velocity to the stellar escape velocity
($v_\infty$/$v_{\rm esc}$) for the SMC stars was used to show that the
`bi-stability jump' in the behaviour of stellar winds at 21,000\,K
(Lamers et al., 1995) is a more gentle transition than previously
thought (Evans et al., 2004b).  Indeed, a more quantitative treatment
of Galactic B-type supergiants found a comparable trend (Crowther et
al., 2006), directly relating this to the distinctive change seen in
the UV morphology between B0.5 and B0.7 subtypes (Walborn \&
Nichols-Bohlin, 1987).  Note that the STIS data have much broader
applications as a metal-poor spectral library, useful 
in the context of disentangling the integrated-light
observations of distant super-star-clusters (e.g. V{\'a}zquez et
al., 2004).  They have also been incorporated in the population synthesis
code {\sc starburst}{\small 99} (Leitherer et al., 2001).

Far-UV, high-resolution spectroscopy from {\it FUSE} provided access
to a wealth of additional diagnostic lines (Walborn et al.,
2002).  In many cases the {\it FUSE} spectra have enabled the first
precise measurements of $v_\infty$ for
stars previously with only {\it HST}-FOS or {\it IUE} data 
(e.g. Crowther et al., 2002; Evans et
al., 2004c), as well as detailed studies of the wind structure and
ionization (Massa et al., 2003).  It also transpires that the P~{\small V} 
and S~{\small IV} lines in the far-UV are sensitive to the
stratification of the winds, providing diagnostics of their
`clumpiness' (Crowther et al., 2002; Evans et al., 2004c; Fullerton
et al., 2006).

\section{Spectral analysis with improved model atmospheres}\label{models}

The continued development of model atmosphere codes has also been a
crucial ingredient to studies of massive stars over the past decade.
The most commonly used non-LTE, line-blanketed codes that take into
account spherical geometry and the effects of the stellar wind are
{\sc cmfgen} (Hillier \& Miller, 1998; Hillier et al.  2003) and {\sc
fastwind} (Santolaya-Rey et al., 1997; Puls et al., 2005); both codes
also include parameterizations for the effects of clumping in the
wind.  Compared to previous results (e.g. Vacca et al., 1996), these
developments led to a downward revision of the temperature scale
for a given spectral type (e.g. Martins et al., 2002; Crowther et al.,
2002; Repolust et al., 2004).

In cases where the winds are less significant, plane-parallel model
atmospheres from {\sc tlusty} (Hubeny \& Lanz, 1995) are also used
widely in the analysis of early-type stars.  Even in the case of
B-type supergiants, which have relatively extended atmospheres, if one
excludes the features most influenced by the stellar wind (e.g. H$\alpha$
and H$\beta$), good agreement was found in the atmospheric parameters
and chemical abundances obtained from {\sc tlusty} models compared
with results obtained with {\sc fastwind} (Dufton et al., 2005).

The combination of new UV data, high-resolution optical spectroscopy
and improved model atmospheres led to a number of
multi-wavelength analyses of individual O-type stars in the Clouds:
Crowther et al. (2002); Hillier et al. (2003); Bouret et al. (2003);
Evans et al. (2004c); Heap et al. (2006).  The wind properties and
chemical abundances of B-type supergiants in the SMC were also
investigated (Trundle et al., 2004; Trundle \& Lennon, 2005).

While these studies began to explore some of the questions posed
regarding stellar evolution in the Clouds, they lacked sufficiently
large samples, in terms of the sampling of spectral types and
luminosities.  The {\sc fastwind} analyses of 40 O-type stars from
Massey et al. (2004, 2005) went some way to address the broader
questions in the SMC compared to Galactic samples (see subsequent
sections), but there remained a strong desire for a large, homogeneous
sample which also included observations of early B-type stars to
investigate the effects of rotation and rotationally-induced mixing.

\newpage
\section{The VLT-FLAMES Survey of Massive Stars}
The delivery of the FLAMES instrument to the VLT in 2002 was the
catalyst for an ESO Large Programme (P.I.  Smartt) to investigate the
role of metallicity in the evolution of massive stars.  Seven
fields centred on stellar clusters were observed: NGC\,3293,
NGC\,4755, and NGC\,6611 in the Galaxy; NGC\,2004 and N11 in the LMC;
NGC\,330 and NGC\,346 in the SMC.  In total, high-resolution
spectroscopy was obtained for $\sim$700 O- and early B-type stars
(Evans et al., 2005; 2006).  All of the observed OB-type stars have
now been analysed to yield physical parameters, chemical compositions and
rotational velocities, as summarised by Evans et al. (2008).  Some of
the key results from the survey are described in the rest of this review, combined
with new results from other studies.

\subsection{Present-day composition of the LMC \& SMC}
Determinations of chemical abundances in rapidly-rotating stars are
complicated by their broadened lines -- in part the reason why
previous observational effort has focussed mostly on narrow-lined
(i.e. slowly-rotating) stars.  To inform analysis of the whole FLAMES
sample, the narrow-lined B-type stars ($v$sin$i <$~100\,kms$^{-1}$) were
analysed first (Hunter et al., 2007; Trundle et al., 2007), yielding
stellar abundances for 87 stars in the Clouds.  The present-day
composition of the Clouds, as traced by these B-type stars is listed in
Table~\ref{tab1} (Mokiem et al., 2007b).  Note that due to
uncertainties in the absolute abundances, the fractions quoted for
iron are relative to the Galactic results from the
FLAMES survey.

\begin{table}[h]
\begin{center}
\vspace*{-0.35cm}
\caption{Present-day composition of the LMC and SMC, as traced by early B-type
stars observed in the FLAMES survey.  Abundances are given on the scale 12$+$log[X/H], 
with the relative fraction compared to the Solar results (Asplund et al., 2005) given in
brackets.}\label{tab1}
\vspace*{-0.1cm}
\begin{tabular}{lccc}
&&&\\
\hline
Element & Solar & LMC & SMC \\
\hline
C & 8.39 & 7.73 & 7.37 \\
& & [0.22] & [0.10] \\
\hline
N & 7.78 & 6.88 & 6.50 \\
& & [0.13] & [0.05] \\
\hline
O & 8.66 & 8.35 & 7.98 \\
& & [0.49] & [0.21] \\
\hline
Mg & 7.53 & 7.06 & 6.72 \\
& & [0.34] & [0.15] \\
\hline
Si & 7.51 & 7.19 & 6.79 \\
& & [0.48] & [0.19] \\
\hline
Fe & 7.45 & 7.23 & 6.93 \\
& & [0.51] & [0.27] \\
\hline
\end{tabular}
\end{center}
\end{table}

\vspace*{-0.05in} Given the strong evolutionary effects on nitrogen
enrichment, it is difficult to obtain the pristine value.  However,
the lowest abundances from the FLAMES results are in good agreement
with estimates from H\,{\small II} regions, leading to their adoption
here (see discussion by Hunter et al., 2007).  The oxygen abundances
are in excellent agreement with results from H\,{\small II} regions,
e.g., 12+log[O/H] = 8.35 and 8.03 for the LMC and SMC, respectively,
from Russell \& Dopita (1992).  It has been known for some time that
the initial abundances of carbon and nitrogen are relatively
more under-abundant than the heavier elements in the Clouds; the
FLAMES results reinforce the varying fractions from element to
element.  Indeed, simply scaling solar abundances for quantitative
work in the Clouds does not best reproduce the observed patterns.

\section{Metallicity-dependent stellar winds}

Analysis of O-type spectra can be a complex, time-consuming process.
In addition to the usual parameters used to characterise a star
(temperature, luminosity, gravity, chemical abundances), we also
need to describe the velocity structure and mass-loss rate of the wind.
The analysis of the FLAMES data used a semi-automated approach,
employing genetic algorithms (GA) to fit the observations with
synthetic spectra from {\sc fastwind} model atmospheres.  This method
was tested using a sample of Galactic stars, finding good agreement with
previous results (Mokiem et al., 2005).  The O-type spectra from the
FLAMES survey were then analysed using the GA approach (Mokiem et al.,
2006; 2007a).

\begin{figure}[h]
\vspace*{-0.25 cm}
\begin{center}
\includegraphics[width=4.75in]{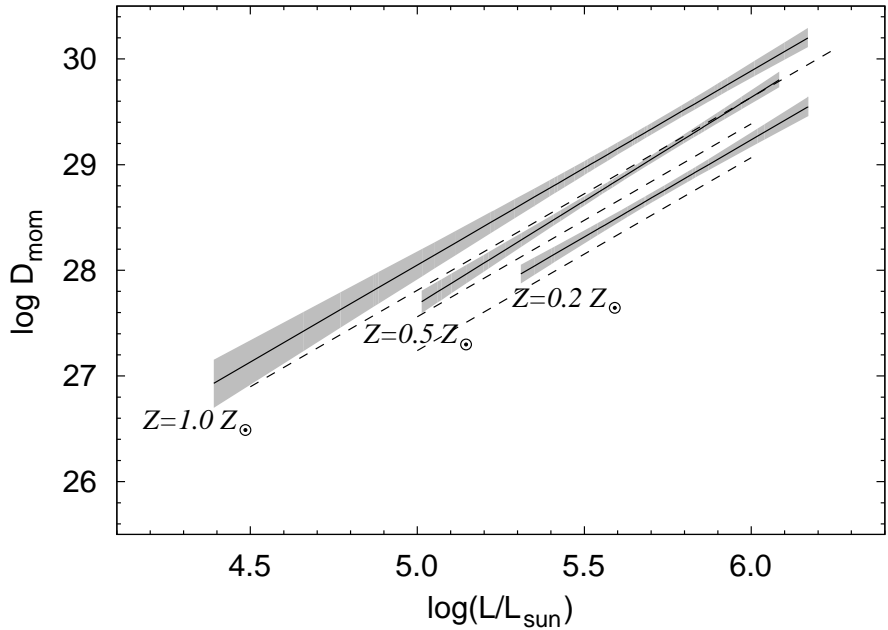} 
\vspace*{-0.1 cm}
\caption{Comparison of the observed wind-momentum--luminosity relations (solid lines)
with theoretical predictions (dashed lines) for O-type stars from \cite{m07b}.  The upper, middle and
lower relations correspond to Galactic, LMC and SMC results, respectively.}\label{fig2}
\end{center}
\end{figure}

The effect of metallicity on wind intensities in O-type stars was
investigated by Massey et al. (2005).  From comparisons of the
wind-momenta (D$_{\rm mom}$, a function of the mass-loss rate,
terminal velocity and stellar radius) for a sample of 22 stars in the
Clouds, Massey et al. found evidence for an offset with $Z$.  This
result is seen more clearly in the FLAMES results, as shown in
Fig.~\ref{fig2}, providing compelling evidence for reduced intensities at
decreased metallicities.  Fig.~\ref{fig2} also shows the theoretical
predictions using the prescription from Vink et al.  (2001).  The
relative separations are in good agreement, with the FLAMES results
finding a scaling of $Z^{0.72-0.83}$ (with the exponent depending on
assumptions regarding clumping in the winds), as compared to
$Z^{0.69\pm0.10}$ from theory (Mokiem et al., 2007b).

\begin{figure}[h]
\vspace*{-0.35 cm}
\begin{center}
\includegraphics[scale=0.53,angle=90]{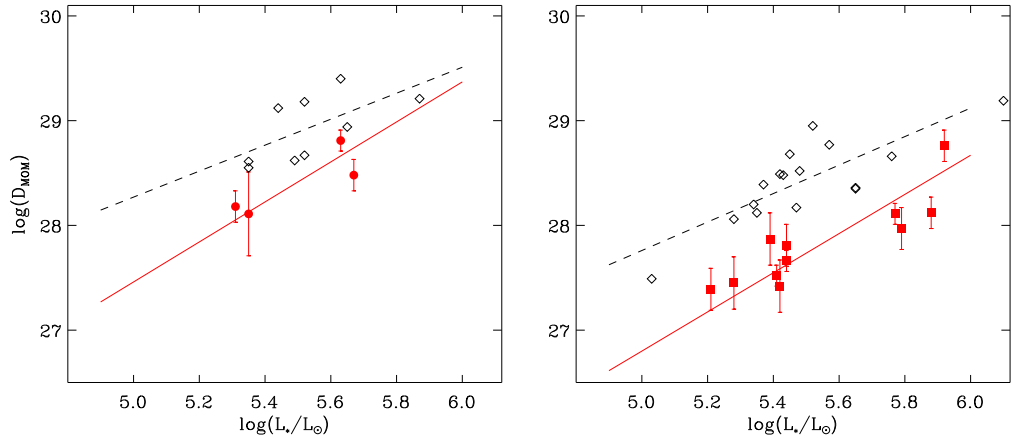} 
\vspace*{-0.1 cm}
\caption{Observed wind-momentum--luminosity relations for early
B-type supergiants ({\it left:} T$_{\rm eff}>$~23.5\,kK;
{\it  right:} T$_{\rm eff}<$~23.5\,kK) in the SMC (solid line,
filled circles/squares) from \cite{tl04} and \cite{tl05}, compared to
Galactic results (dashed line, open diamonds) from \cite{clw06}.
Figure from \cite{ct_tartu}.}\label{fig3}
\end{center}
\end{figure}

There is also quantitative evidence for weaker winds in early B-type
supergiants in the SMC compared to their Galactic counterparts --
somewhat reassuring given that they are the direct descendants
of massive O-type stars!  Fig.~\ref{fig3} shows the SMC results from
Trundle et al. (2004) and Trundle \& Lennon (2005), compared to 
Galactic results from Crowther et al. (2006).

These observational tests are important for a number of areas --
including considerations of the feedback from massive stars to the
local interstellar medium in the context of star formation (see review
by Oliveira, these proceedings).  The reduced mass-loss rates at lower
$Z$ mean that less angular momentum is lost, i.e. an evolved star in
the SMC would be expected to retain a larger fraction of its initial
rotational velocity compared to a similar star in the Galaxy.  Indeed,
the rotational velocity distribution for the unevolved
(i.e. luminosity class IV or V) SMC stars, appears to have
preferentially faster velocities when compared to Galactic results
(albeit limited in terms of its statistical significance; Mokiem et
al., 2006).  This could offer a potential channel for long duration,
gamma-ray bursts at low metallicity (e.g. Yoon et al. 2006).

Reduced mass-loss also impacts on the end products of massive O-type
stars.  As summarised by Crowther (2007), in the `Conti Scenario'
stars in the Milky Way can expect to pass through the following phases,
depending on their initial masses (M${\rm _i}$):
\begin{itemize}
\vspace*{0.175cm}
\item{M$_{\rm i} >$75 M$_\odot$: O $\rightarrow$ WN(H-rich) $\rightarrow$ LBV $\rightarrow$ WN(H-poor) $\rightarrow$ WC $\rightarrow$ SNIc;}
\vspace*{0.175cm}
\item{M$_{\rm i}$ = 40-75 M$_\odot$: O $\rightarrow$ LBV $\rightarrow$ WN(H-poor) $\rightarrow$ WC $\rightarrow$ SNIc;}
\vspace*{0.175cm}
\item{M$_{\rm i}$ = 25-40 M$_\odot$: O $\rightarrow$ LBV/RSG $\rightarrow$ WN(H-poor) $\rightarrow$ SNIb.}
\vspace*{0.175cm}
\end{itemize}

With reduced mass-loss, the threshold to reach the WC phase in the
SMC will move upwards from 40\,M$_\odot$.  This is reflected in the
small relative number of WC to WN stars at low metallicity, as
compared to the ratio seen in, e.g. the solar neighbourhood and M31
(see Fig.~8 from Crowther, 2007).  

Moreover, a clearer picture is emerging
of the progenitors of core-collapse supernovae, with increasing
evidence that type II-P supernovae are from red supergiants (see 
discussion by Smartt et al., 2008, and references therein).
As such, one would expect that the ratio of type II to type Ib/c
supernovae also varies with $Z$, with initial evidence of this recently
reported by Prieto et al. (2008).

\section{The stellar temperature scale as a function of metallicity}

\subsection{O- and early B-type stars}
\begin{figure}[h]
\vspace*{-0.35 cm}
\begin{center}
\includegraphics[width=4in]{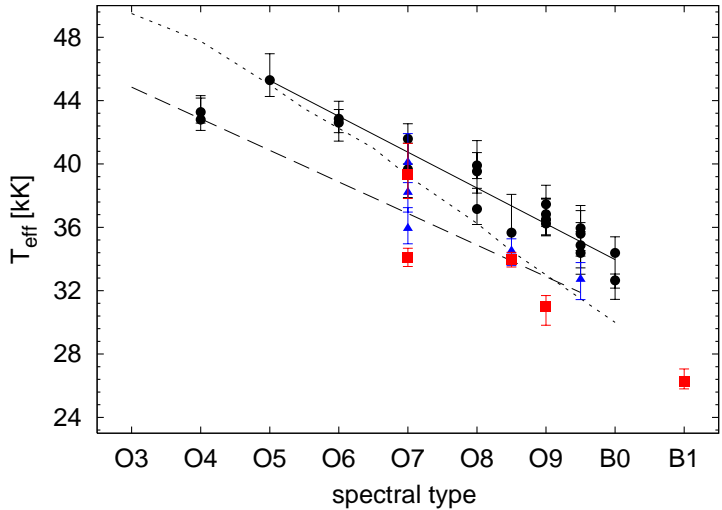}
\vspace*{-0.1 cm}
\caption{Effective temperatures for the O-type SMC stars analysed
by Mokiem et al. (2006).  The solid line is the fit to the SMC dwarfs, compared to the 
calibration for Galactic dwarfs from Martins et al. (2005).  The dotted line is the 
SMC scale from Massey et al. (2005) for luminosity class V and III stars.}\label{fig4}
\end{center}
\end{figure}

Stellar temperatures have long been known to depend on luminosity
class -- supergiants, with their lower gravities and extended
atmospheres, are found to be cooler than dwarfs of the same spectral
type.  The inclusion of line-blanketing effects in the calculation of
model atmospheres led to an overall downward revision of the stellar
temperature scale (see Section~\ref{models}).  However, this effect is
less dramatic at lower metallicities because of the diminished
cumulative opacity from the metal lines -- there is less `back
warming' by trapped radiation, and so a hotter model is required to
reproduce the observed spectral line-ratios (ionization balance).
This effect is clearly seen in the temperatures obtained for O-type
stars in the SMC, which are hotter than those found for Galactic stars
with the same spectral type (Massey et al., 2005; Mokiem et al. 2006;
Fig.~\ref{fig4}).  The temperatures for stars in the LMC are seen to fall neatly between
the SMC and Galactic results (Mokiem et al., 2007a).  A similar
$Z$-dependence has also been seen for the first time in the 
early B-type stars observed with FLAMES (Trundle et al., 2007).

\vspace*{-0.35cm}
\subsection{Late-type supergiants}
The classification of O- and early B-type stars is based, primarily, on the
relative line-ratios of different ionization stages of the same
element (helium and also silicon).  The $Z$-dependence of
the temperature scale is a consequence of the abundance effects in the
model atmosphere calculations.  

\begin{figure}[t]
\vspace*{-0.35 cm}
\begin{center}
\includegraphics[width=4in]{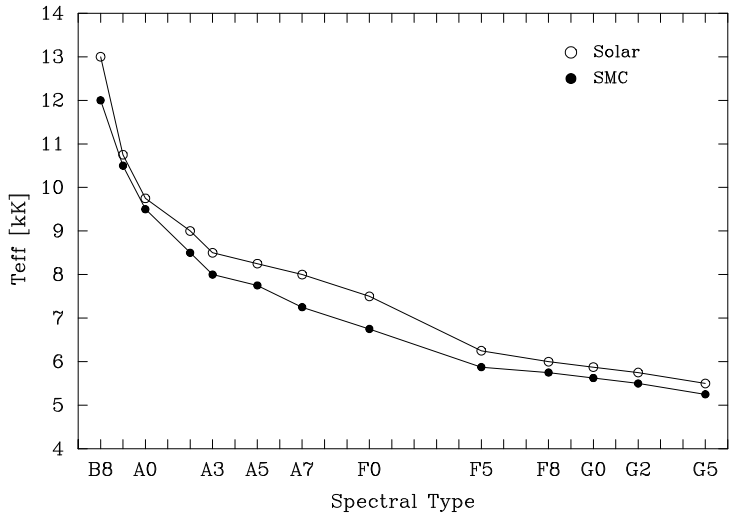}
\vspace*{-0.1 cm}
\caption{Temperature calibrations for late-type supergiants (Evans \& Howarth, 2003).}\label{fig4b}
\end{center}
\end{figure}

For evolved, luminous supergiants there is a more direct effect of the
metal abundances on the stellar temperature scale.  For
instance, the primary classification criterion in the A-type domain is
the ratio of the Ca~$K$ line to the blend of the Ca~$H$ and the
H$\epsilon$ line.  Thus, a cooler temperature is needed to reproduce
the criterion for a given spectral type at SMC metallicity than in a
Galactic star.  This effect is the origin of the cooler temperatures
reported by Venn (1999), which led her to reclassify those stars.
However, the classifications should be employed on purely
morphological grounds, i.e. T$_{\rm eff}$ can be {\it f(Z)} at a given
spectral type, but the spectral type should be independent of
environment.  Evans \& Howarth (2003) investigated the scale of this
effect, finding lower temperatures in the SMC by up to 10\%
(Fig.~\ref{fig4b}).  This effect, although less significant, is also
seen in F- and G-type supergiants.

Temperatures determined recently for M-type supergiants in the Galaxy
and the Clouds also find a $Z$-dependence, again with the SMC stars
cooler at a given type (Levesque et al., 2005; 2006; 2007).  The
primary classification criterion in this domain is the intensity of
the TiO bands, so cooler temperatures are also required to yield the
intensity necessary for a given spectral type -- an effect first
noted by Humphreys (1979), while attempting to explain
the later types seen in M-type supergiants in the Milky Way and LMC
compared to the SMC.  A similar trend in temperatures is
also seen between the LMC and SMC results for K-type supergiants
although, somewhat intriguingly, the Galactic stars appear slightly
cooler at the early K types (Fig.~6, Levesque et al., 2007).

In summary, effective temperatures (for a given spectral type) are a
function of metallicity over the whole of the upper
Hertzsprung-Russell diagram -- from the most massive O-type stars,
through to the coolest M-type supergiants.  The typical variations
between solar and SMC metallicity are relatively small ($\sim$5-10\%),
but should be considered when adopting temperature estimates on the
basis of a known spectral type.

\vspace*{-0.3cm}
\section{Evidence for faster rotation at low $Z$}
The first large study of the $Z$-dependence of stellar rotational
velocities was the work by Keller (2004), who observed $\sim$100
B-type stars in the LMC and compared their rotation rates with
published Galactic results.  Keller found that the cluster members had
mean rotational velocities that were larger than field stars (at the
corresponding metallicity), and that the LMC cluster members were
rotating more quickly than those in Galactic clusters (at just under a
2$\sigma$ significance).  Since then, a number of surveys have used
FLAMES to investigate these effects further in early B-type stars (in
which the effects of mass-loss are relatively small, thereby removing
a further complication to the evolution of $v$sin$i$).

\begin{figure}[t]
\vspace*{-0.35 cm}
\begin{center}
\includegraphics[width=4.5in]{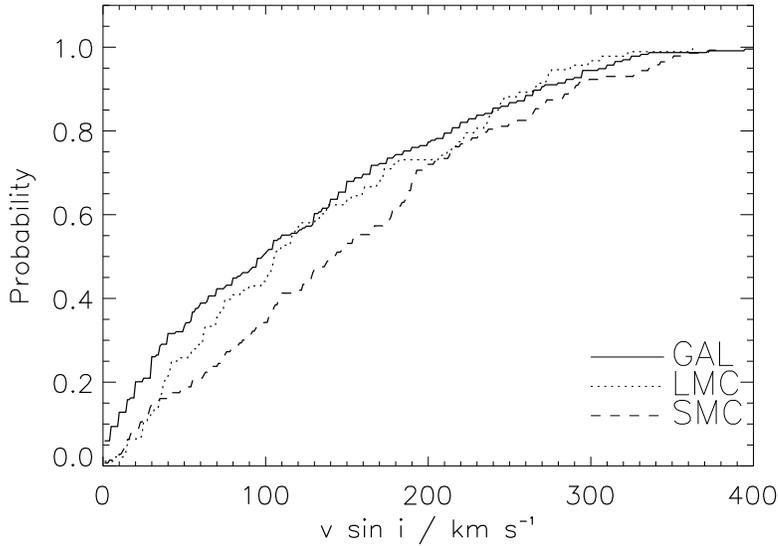} 
\vspace*{-0.1 cm}
\caption{Cumulative distribution functions for the rotational velocities of Galactic
field stars, compared with LMC and SMC results from FLAMES (Hunter et al., 2008a).}\label{fig5}
\end{center}
\end{figure}

From FLAMES observations of B- and Be-type stars in fields
centred on NGC\,330 and NGC\,2004, Martayan et al. (2007) found
evidence for faster rotational velocities at the lower metallicity of the SMC.
Similar conclusions were found from the analysis of Hunter et
al. (2008a; Fig.~\ref{fig5}), in which the SMC stars are
found to rotate more quickly than those in the Galaxy, with a
significance at the 3$\sigma$ level.  Given the overlapping target
fields, an independent check on the methods was possible between the
results of Martayan et al. and Hunter et al., finding good agreement
in the velocity distributions of normal B-type stars in the NGC\,330
field.  

To complicate this picture slightly, new results from additional
FLAMES observations (Royer et al., this meeting ) find velocity
distributions in the LMC and SMC that are consistent with being drawn
from the same parent population -- this raises the question of whether
local metallicity variations could account for the differences?  Given
the large global offset between the LMC and SMC, this seems relatively
unlikely, but it appears that something is still eluding us at the
current time!

\vspace*{-0.1in}
\section{Reconciling the `Hunter diagram' with evolutionary predictions}

\begin{figure}[h]
\begin{center}
\hspace*{-0.55cm}\includegraphics[width=5.6in]{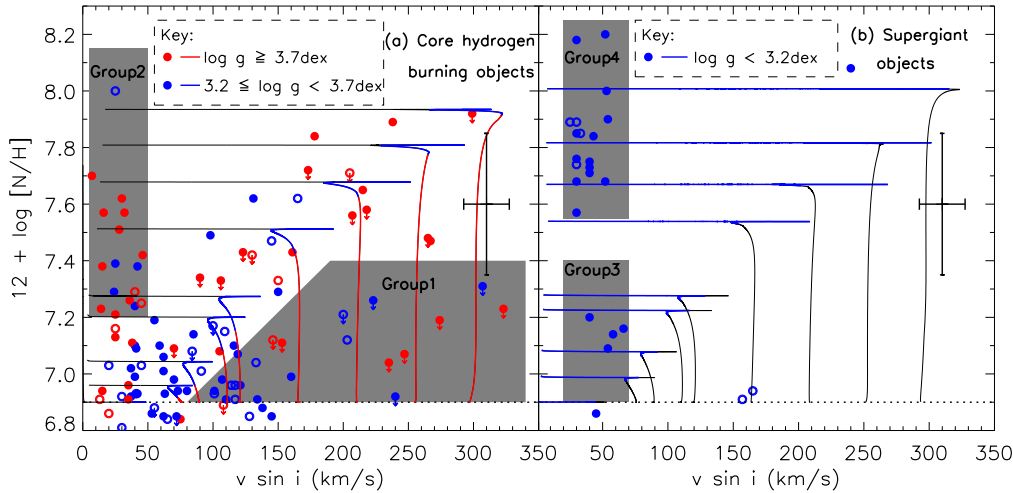} 
 \caption{Nitrogen abundances (12$+$log[N/H]) compared to projected rotational
velocities for core-hydrogen-burning (left-hand panel) and supergiant (right-hand
panel) B-type stars in the LMC (Hunter et al., 2008b).  The solid lines are new
evolutionary tracks, open circles are radial velocity variables, downward arrows are
upper limits, and the dotted horizontal line is the LMC baseline nitrogen abundance.}\label{fig6}
\end{center}
\end{figure}

To investigate the impact of rotation on surface nitrogen abundances
in the FLAMES sample, new evolutionary models were calculated using
the chemical compositions from Table~1.  Fig.~\ref{fig6} shows the
nitrogen abundances, as a function of $v$sin$i$ for the B-type
stars in the LMC from the FLAMES survey (Hunter et al., 2008b).  Typical
uncertainties in the abundances are $\sim$0.2-0.3 dex, so the scatter
of the results indicates genuine differences in the N-enrichment in
both the core-hydrogen-burning stars (dwarfs and giants, left-hand
panel) and the supergiants (right-hand panel).

Two groups (`Groups 1 and 2') appear inconsistent with the predicted
abundances.  The dark grey (blue) points in Group 1 comprise
rapidly-rotating stars that appear to have undergone little chemical
mixing, and yet they have surface gravities that indicate they are
near the end of core hydrogen burning.  These results are at odds with
the evolutionary models that predict nitrogen abundances some
$\sim$0.5 dex greater for the more massive stars (in which mixing is
expected to be most efficient).  Note that there was no evidence for
binarity in the spectra of many of these stars.

The 14 (apparently single) core-hydrogen-burning stars in Group 2 are
equally puzzling as they are rotating very slowly ($v$sin$i<$50 kms$^{-1}$),
but show significant N-enrichment.  For a random orientation, we
could expect about two of these to be rapidly-rotating stars viewed
pole-on.  This seems an unlikely explanation for all 14, and
Hunter et al. concluded that the majority are intrinsically slow
rotators.  Recent studies of Galactic $\beta$-Cepheid stars have found
a correlation between nitrogen enrichment and magnetic fields (Morel
et al., 2006); perhaps the enrichments in Group 2 are related to
magnetic fields.

The results for the supergiants can be considered as two groups --
Group 3, with relatively normal levels of N-enrichment, and Group 4,
with much larger abundances.  Simplistically these could be pre-RSG
(Group 3) and post-RSG (Group 4) stars.  However, while the abundances
of the Group 4 stars are consistent with the predicted level, the
models cannot reproduce their effective temperature in the
Hertzsprung-Russell diagram.  Some of these show evidence of binarity,
so mass transfer may also be important.

So, rotationally-induced mixing appears to play a key role in the enrichment
of surface nitrogen in massive stars, but it appears that there are also other
processes at work, particularly at low rotational velocities -- presenting new
challenges to the evolutionary models.

\vspace*{-0.2in}
\section{Massive binaries}
The effects of binarity/multiplicity on the formation and subsequent
evolution of high-mass stars are a vibrant area of research.  One of
the key ingredients missing from current theories of both star
formation and cluster evolution is a robust binary fraction of
high-mass stars (and the distribution of the relative mass-ratios in
these systems).  Observational effort in this area has been somewhat
piecemeal to date (e.g. Sana et al., 2008), with Zinnecker \& Yorke
(2007) highlighting the need for multi-epoch radial velocity surveys of
stellar clusters to provide better constraints to theoretical models
of star formation.

One of the serendipitous aspects of the FLAMES survey was the large
number of spectroscopic binaries discovered.  The time sampling of the
service-mode observations did a reasonable (but not thorough) job of
binary detection in three of the fields in the Clouds, with lower
limits to the binary fraction of 25-35\% (Evans et al., 2006).  As
illustrated by new multi-epoch observations in 30~Doradus with GMOS
(Bosch et al., these proceedings), the true binary fraction in young clusters
could be much larger.

\vspace*{-0.1in}
\section{Closing remarks}

There has been huge observational and theoretical progress in our
understanding of massive star evolution in the Clouds in the past
decade.  The intensity of stellar winds, the effective temperature
scale, and the rotational velocities of OB-type stars are all found to
be dependent on the metal content of their local environment.

Multi-object spectrographs (e.g. FLAMES, 2dF and GMOS) have truly
opened-up our knowledge of the stellar content of the Clouds over the
past decade.  We are beginning to discover many new examples of the
peculiar members of the `OB Zoo' (e.g. Vz, f?p, nfp, B[e], etc.) and
with future surveys we will be able to explore the
evolutionary connection of these rare sub-types to the morphologically
normal population -- are they critical, short-lived phases that
every massive star experiences?

Moreover, key questions remain regarding the binary fraction of
massive stars, binary evolution, and the efficiency of
rotationally-induced mixing in O-type stars.  The VLT-FLAMES Tarantula
Survey is an ESO Large Programme (P.I. Evans) approved in July 2008 to
address these issues, via multi-epoch observations of $\sim$1,000
stars in 30~Doradus in the LMC.

Lastly, it is worth noting that {\it IUE} left a large spectral
archive of Galactic OB-type stars, which has continued to yield new
results.  While {\it FUSE} has observed a large number of early-type
stars (giving us a valuable window on the wind parameters revealed in
the far-UV), it is remarkable that high-resolution
spectroscopy from {\it HST} exists for only $\sim$40 stars in the
Clouds.  The 1200-1900\AA\/ region is uniquely important in the study
of massive stars and, given the absence of future UV missions at the
advanced stages, new {\it HST} observations after the servicing
mission would provide immense legacy value.

\acknowledgements{Grateful thanks to Nolan Walborn for careful reading
of this manuscript.}

\end{document}